\documentclass[preprint,12pt]{elsarticle}

\usepackage{amssymb,amsmath,bm,nicefrac,array,color,float,subfigure,dsfont,txfonts,wasysym,multirow,sidecap,xcolor,cancel,dcolumn}

\usepackage[plainpages=false,pdfpagelabels,
    colorlinks=true,linkcolor=red,urlcolor=blue,
    citecolor=blue,pdftitle={Cr-SC-v1},pdfauthor={},
    pdfdisplaydoctitle=true]{hyperref}
\usepackage{cleveref}

\begin{document}

\begin{frontmatter}

\title{Doping-dependent orbital magnetism in Chromium pnictides}

\author{Henri~G.~Mendon\c{c}a}
\affiliation{organization={Instituto de F\'isica, Universidade Federal de Uberl\^andia}, 
city={Uberl\^andia}, state={Minas Gerais}, postcode={38400-902}, country={Brazil}}

\author{George B. Martins}
\affiliation{organization={Instituto de F\'isica, Universidade Federal de Uberl\^andia}, 
city={Uberl\^andia}, state={Minas Gerais}, postcode={38400-902}, country={Brazil}}

\author{Lauro B. Braz}
\affiliation{
organization={Instituto de F\'{\i}sica, Universidade de S\~ao Paulo}, city={S\~ao Paulo}, state={S\~ao Paulo}, 
postcode={05508-090}, country={Brazil}
}

\begin{abstract}
	We present results for the phase diagram of the parent compound LaCrAsO under electron doping using the matrix random-phase approximation. 
    At low doping levels, the system stabilizes an antiferromagnetic state in which different Cr sublattices carry opposite spins, consistent with experimental observations. 
    As the doping concentration increases, a stripe-type antiferromagnetic phase becomes favored. 
    At even higher doping, the system repeats the two former magnetic states, but with incommensurate magnetic ordering vectors. 
    The commensurate magnetic phases are associated with more localized electrons in the Cr $d_{3z^2-r^2}$ orbital, whereas the incommensurate phases are linked to the $d_{xy}$ orbital, whose stronger overlap favors itinerant-electron magnetism.
\end{abstract}

\end{frontmatter}

\section{Introduction} 

The phase diagram of cuprate superconductors has, since the early days, clearly shown that 
electron correlations decrease with hole doping (HD), going from the pseudo-gap strange-metal 
underdoped region to the normal-metal overdoped region, passing through the intermediate 
optimal-doping maximum-${\rm T_c}$ region \cite{Lee2006}. Something similar occurs on the electron doping (ED) side 
of the phase diagram, despite some differences in relation to the HD side: the ED side has lower ${\rm T_c}$'s, 
the superconducting (SC) phase spans a much narrower doping range, and 
a spin-density-wave description of the normal metallic state (near optimal doping), treated at the mean-field level, 
offers a good description of quite a few experimental results (for details, see Armitage \emph{et al.}~\cite{Armitage2010}). 
One of the possible explanations for these differences is that correlations are weaker in the ED side. 
Sitting in the middle of the HD and ED regions is a half-filling (zero-doping) antiferromagnetic (AF) Mott-insulator. 
The many similarities between cuprates and Iron-superconductors 
(Fe-SC) have lead, since the discovery of the latter, to efforts to find a way of placing under the same umbrella the 
phase diagrams of both materials. An interesting work by Ishida and Liebisch~\cite{Ishida2010} pointed out that 
if one defines the Fe-SC doping concentration ($\delta=\nicefrac{n}{5}-1$) not in relation to 
$n=6$ (Fe$^{2+}$), but in relation to the half-filled $n=5$ configuration, in the range $0 \le \delta \lesssim 0.4$ 
Fe-SC and cuprates display the same sequence of phases, namely, Mott-insulator/non-Fermi-liquid (bad-metal)/Fermi-liquid, 
which occur roughly at the same concentration intervals~\cite{note1}, with the parent compound LaFeAsO ($\delta=0.2$)
sitting at the boundary between the non-Fermi-liquid and Fermi-liquid phases. 

The insight described in Ref.~\cite{Ishida2010} rests on a comparison 
of ED for Fe-SC with HD for the cuprates. As pointed out in 
Refs.~\cite{Pizarro2016,Edelmann2017,Wang2017}, if one presses the analogy further, it is expected that 
a possible SC phase obtained by \emph{hole-doping} a putative Fe-SC Mott-insulating phase (i.e., starting from 
$n=5 \equiv \delta=0$) may result in higher $T_c$ values, as is the case for the cuprates \cite{note2}. 
Given the large number (5) of active $3d$ orbitals in the Fe-SC, a wide range of variation in $\delta=\nicefrac{n}{5}-1$ 
requires a large variation in $n$. In addition, as the Fe-SC parent compounds already start in the electron doped side 
of the Mott insulator ($n=6 \equiv \delta=0.2$), looking for parent compounds with lower values of $n$ would be a natural 
step in reaching the hole-doped side of the phase diagram. The transition metal ions to consider would then be Manganese and Chromiun. 
In that respect, ${\rm BaMn_2As_2}$, a compound isostructural to ${\rm BaFe_2As_2}$, 
is an AF insulator (checkerboard G-type order, $T_N=625$~K \cite{Singh2009a,Singh2009b,PhysRevMaterials.7.044410}), or rather a small band-gap 
semiconductor with activation energy $\approx 0.03$~eV \cite{Singh2009a}, where an insulator-to-metal 
transition occurs with hole doping via the substitution of a few percent of Ba with K \cite{Bao2012,Pandey2012,PhysRevB.97.014402}. 
This transition causes only a small suppression of $T_N$ and of the ordered magnetic 
moment (which varies between $S \sim 2$ to $\nicefrac{5}{2})$, suggesting that doped holes interact weakly with 
the Mn spin system, and therefore, up to now, no superconductivity has been reported in this 
compound. It has also been found that ${\rm BaMn_2As_2}$ undergoes an insulator-to-metal 
transition under pressure of $\sim 4.5$~GPa \cite{Satya2011,PhysRevMaterials.7.044410}. Attempts to grow ${\rm Ba(Mn_{1-x}T_x)_2As_2}$ 
(T = Cr, Fe, Co, Ni, Cu, Ru, Rh, Pd, Re, and Pt) and ${\rm BaMn_2(As_{1-x}Sb_x)_2}$ were mostly unsuccessful, 
as the substitution values for the majority of these compounds does not exceed $0.5\%$ \cite{Pandey2011}. 
Pandey \emph{et al.} \cite{Pandey2012} demonstrated that ${\rm Ba_{1-x}K_xMn_2As_2}$ 
becomes a metal with partial substitution of K for Ba, while retaining
the crystal and AF structures of ${\rm BaMn_2As_2}$ and 
with nearly the same high $T_N=625$~K and large ordered moment $\mu=3.9 \mu_B/{\rm Mn}$. 
In reality, Lamsal \emph{et al.} \cite{Lamsal2013} have shown that the local moment 
AF order in ${\rm Ba_{1-x}K_xMn_2As_2}$ remains robust for doping concentrations up to 
$x=0.4$. The chemical potential $\mu$ is nearly independent of $x$ for $0\le x \le 0.4$, 
while $T_N$ decreases to $480$~K for $x=0.4$. 

As to Cr compounds that are isostructural to 122 Fe-SC, we have ${\rm BaCr_2As_2}$, reported to be 
an itinerant (metallic) antiferromagnet \cite{Singh2009,PhysRevB.111.134438}, where DFT calculations have suggested 
a stronger degree of covalency between Cr-As than between Fe-As in Fe-SC's, maybe explaining why 
Cr doping of ${\rm BaFe_2As_2}$ does not result in superconductivity. ${\rm EuCr_2As_2}$ is a metallic 
compound \cite{Paramanik2014}, in which Nandi \emph{et al.} \cite{Nandi2016} 
found, through neutron diffraction, that the ${\rm Cr^{2+}}$ magnetic moment ($1.7 \mu_B$) have AF G-type order 
along the c-axis at a very high transition temperature ($T_N \sim 680$~K). In addition, the ${\rm Eu^{2+}}$ ions are found 
to order ferromagnetically at ${\rm T_C=21}$~K \cite{Nandi2016}, with a negative magnetoristance observed 
below ${\rm T_C}$. The independent magnetic orders in the Cr and Eu layers indicate that they are completely 
decoupled. 

\begin{figure}
\includegraphics[width=1.0\columnwidth]{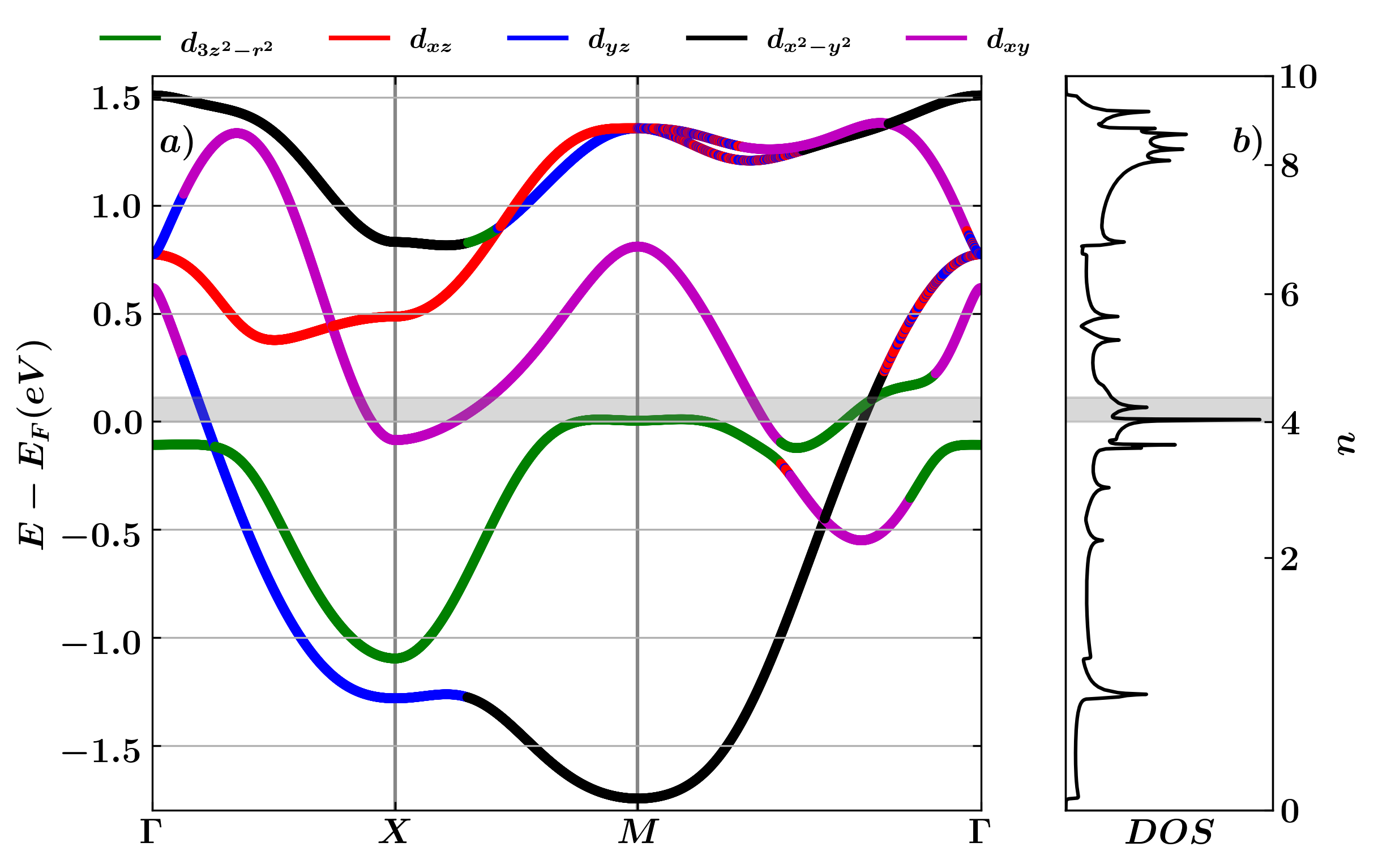}
\caption{(a) Projected band structure in the $d-$orbitals ($d_{3z^2-r^2}$ in green; $d_{xz}$ in red; $d_{yz}$ in blue; $d_{x^2-y^2}$ in black; $d_{xy}$ in magenta) of the tight-binding ${\rm Cr}$ atoms for ${\rm LaCrAsO}$ along the symmetry lines $\Gamma-X-M-\Gamma$. (b) Density of states (DOS) for ${\rm LaCrAsO}$. In panels (a) and (b), the grey regions indicate the doping studied in this work ($n = 4.0 - 4.55$).
}
\label{fig:model}
\end{figure}

In this work, we present multi-band matrix random-phase approximation calculations for the phase diagram of LaCrAsO as a function of electron doping $n$ in the Cr planes. For doping levels between $n=4$ (undoped) and $n \approx 4.2$, the system stabilizes an antiferromagnetic (AF) phase in which the Cr sublattices carry opposite spins. At higher doping, $4.2 \lesssim n \lesssim 4.35$, a stripe-type AF phase is favored, while further doping induces incommensurate states: AF in the range $4.35 \lesssim n \lesssim 4.5$ and stripe-type AF for $n \gtrsim 4.5$.

Our results reveal two central features of this phase diagram. First, the AF phases with opposite Cr sublattice spins are driven by Fermi-surface changes (Lifshitz transitions). Second, the orbital character of magnetism evolves with doping: at low doping ($4 \lesssim n \lesssim 4.25$), it is dominated by the localized $d_{3z^2-r^2}$ orbital, whereas at higher doping ($4.25 \lesssim n \lesssim 4.5$), the planar $d_{xy}$ orbital prevails, marking the crossover from local-moment magnetism to itinerant-electron magnetism.


\section{Theory}\label{sec:model}
\subsection{Band structure}
The tight-binding electronic Hamiltonian $H_{\rm TB}$ describing the low-energy bands of ${\rm LaCrAsO}$ is given by \cite{Wang2017}
\begin{align}
H_{\rm TB} &=\sum_{\mathbf{ k}\sigma\mu\nu} T^{\mu\nu}(\mathbf{ k})
d^\dagger_{\mathbf{ k}\mu\sigma} d^{\phantom{\dagger}}_{\mathbf{ k}\nu\sigma}~, &
\label{eq:hamtb}
\end{align}
where $d^{\phantom{\dagger}}_{\mathbf{ k}\nu\sigma}$ creates an electron with 
momentum $\mathbf{ k}$, spin $\sigma$, in an orbital $\nu$ that is either ${ t_{2g}}$ (${ xy}$, ${ zx}$, 
${ yz}$) or ${ e_g}$ (${ 3z^2-r^2}$ and ${ x^2-y^2}$). The hopping integrals 
entering in the expressions for $T^{\mu\nu}(\mathbf{ k})$ can be found in Ref.~\cite{Wang2017}. 
They were obtained through a standard downfolding procedure from a full DFT 
band structure calculation onto a single-chromium tight-binding model~\cite{Wang2017}.

\begin{figure}
\includegraphics[width=1.0\columnwidth]{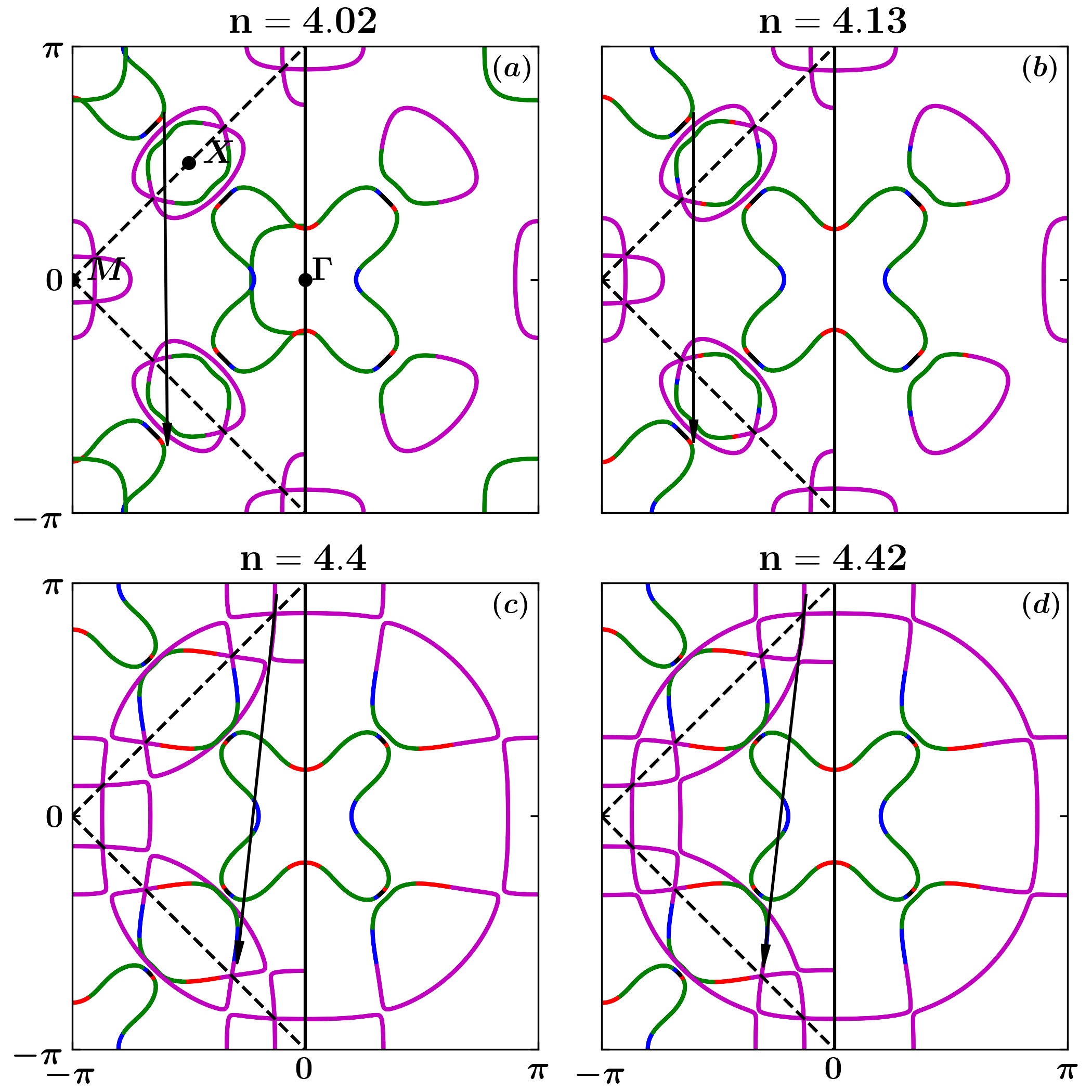}
	\caption{Panels (a) to (d) shows the contribution of the $d-$orbitals ($d_{3z^2-r^2}$ in green; $d_{xz}$ in red; $d_{yz}$ in blue; $d_{x^2-y^2}$ in black; $d_{xy}$ in magenta) to the Fermi surface pockets for ${\rm LaCrAsO}$ for $n = 4.02, 4.13, 4.40,4.42$, respectively. For ${\rm LaCrAsO}$ and in this doping range, the Lifshitz transition occurs between panels (a) and (b), and panels (c) and (d).
    The nesting vectors are represented by the grey arrows on each of the panels.
}
\label{fig:orbital}
\end{figure}

The band structure obtained from the model of Eq.~\eqref{eq:hamtb} is shown in Fig.~\ref{fig:model}$a)$, where colors represent the main orbital contributions at each momentum and energy level.
Along some high-symmetry directions, the orbital contributions become nearly degenerate and then colors mix in Fig.~\ref{fig:model}$a)$.
Panel $b)$ shows the respective density of states along the full energy resolution of the Wannier model.
We highlight the two peaks in the density of states at electron occupations $n\approx4.44$ and $n\approx4.11$.
These peaks are related to two distinct changes in the Fermi surface topology, also known as a Lifshitz transition.
The two transitions are represented on the right-hand side of the panels in Fig.~\ref{fig:orbital}.
Panels $a)$ and $b)$ represent the first Lifshitz transition, in which an electron pocket around $(\pi,\pi)$ is lost.
The second Lifshitz transition is represented in panels $c)$ and $d)$, where pockets around $(\pi,0)$ and $(\pi/2,\pi/2)$ are turned into large pockets centered at $(0,0)$ and $(\pi,\pi)$.

Eq.~\eqref{eq:hamtb} represents a simplified five-band model for the five $d$ orbitals of Cr.
However, in LaCrAsO there is another, inequivalent sublattice site in the square lattice of Cr atoms, which is mapped onto the present Cr $d$-orbital model in real space by a $C_4$ rotation along the $z$ axis and a $(\pi,\pi)$ translation in momentum space.
This procedure is also called band folding, and it was used to obtain the Fermi surfaces shown on the left-hand side of each panel in Fig.~\ref{fig:orbital}.
We remark that the first Brillouin zone is changed by this operation, which is shown by black dashed lines.
The folded Brillouin zone is related to the unfolded one by a $C_4$ operation followed by a rescaling of $1/\sqrt{2}$.

The total Hamiltonian $H_{\mathrm{T}} = H_{\mathrm{TB}} + H_{\mathrm{MB}}$ includes, besides $H_{\mathrm{TB}}$, also 
the many-body term $H_{\mathrm{MB}}$: 
\begin{equation}\begin{split}  \label{eq:Hcoul}
  H_{\rm MB}& =
  U\sum_{{\bf i}\mu}n_{{\bf i}\mu\uparrow}n_{{\bf i}
    \mu\downarrow}
  +U^{\prime}\sum_{{\bf i} \mu < \nu}n_{{\bf i}\mu}n_{{\bf i}\nu}\\
  &\quad +J\sum_{{\bf i}\mu < \nu} (d^{\dagger}_{{\bf i}\mu\sigma}
	d^{\dagger}_{{\bf i}\mu\sigma^{\prime}}d^{\phantom{\dagger}}_{{\bf i}\nu\sigma^{\prime}}
  d^{\phantom{\dagger}}_{{\bf i}\nu\sigma}+h.c.)\\
  &\quad +J^{\prime}\sum_{{\bf i}\mu < \nu}(d^{\dagger}_{{\bf i}\mu\uparrow}
  d^{\dagger}_{{\bf i}\mu\downarrow}d^{\phantom{\dagger}}_{{\bf i}\nu\downarrow}
  d^{\phantom{\dagger}}_{{\bf i}\nu\uparrow}+h.c.),
\end{split}\end{equation}
where $\mu,\nu$ denote the Cr $3d$ orbitals,  
$n_{{\bf i},\nu}$ is the electron density of orbital $\nu$ at Chromiun site
${\bf i}$. The notation is standard and has been described in detail 
in many publications (see for example Ref.~\cite{Kubo2007}). We will 
use the usual relations $U^{\prime}=U-2J$, as well as $J=J^{\prime}$. 
We are then left with two parameters $U$ and $J$. Following standard procedure, 
we will fix the ratio $\nicefrac{J}{U}=$, using $J=\nicefrac{U}{4}$, and thus only 
$U$ will be left free to vary. 
Note that the values for the many-body interactions can be substantially smaller
than the atomic ones, as they may be screened by
bands not included in the Hamiltonian. These many-body terms 
have been discussed in the literature~\cite{Graser2009,Kubo2007}, where more details can be found. 
Energies are given in electron volts.

\begin{figure}
\includegraphics[width=1.0\columnwidth]{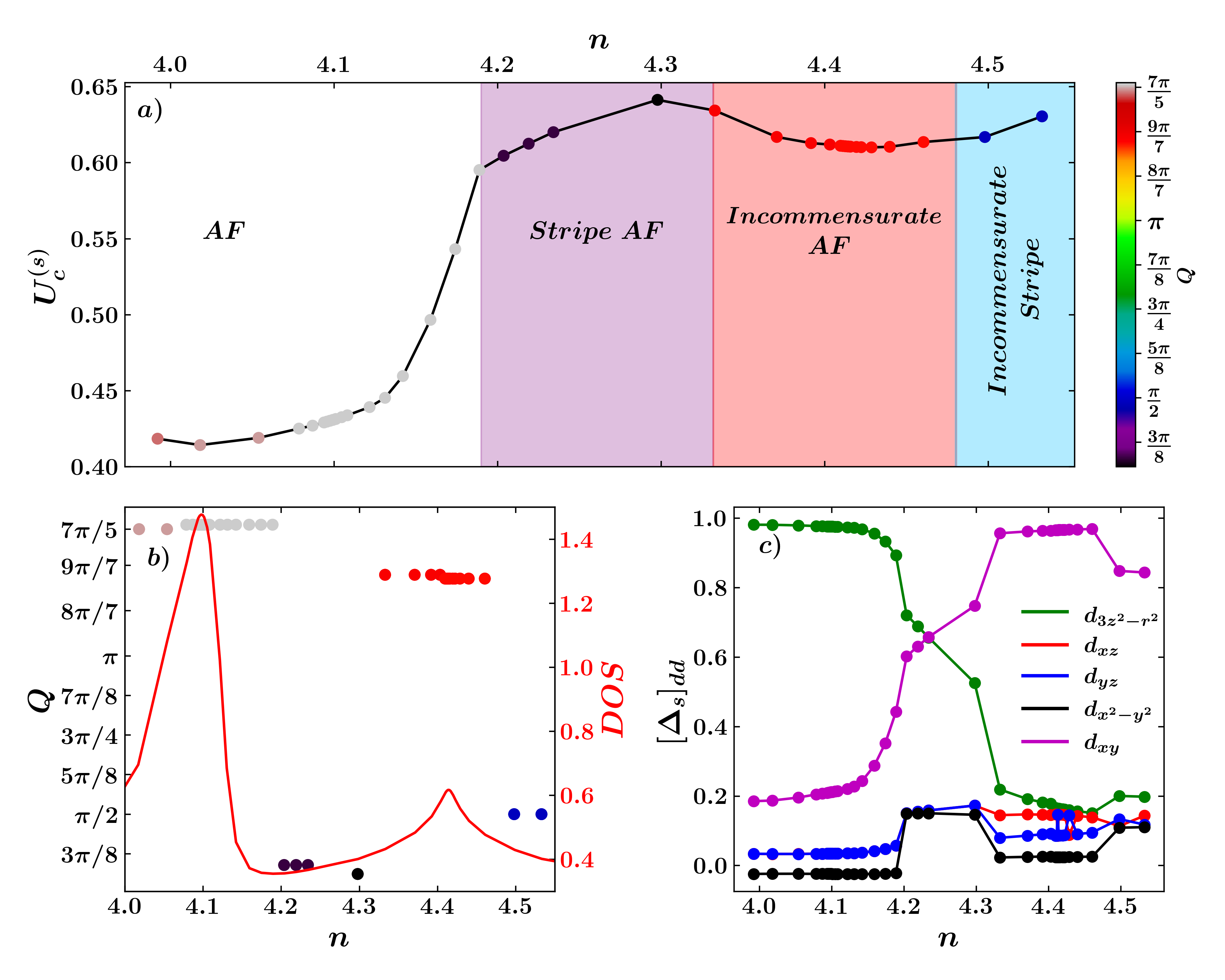}
	\caption{Panels $(a)$ shows the critical Hubbard interaction strength $U_c^{(s)}$ in the spin channel as a function of electron doping $n$. Colors in the scattered points denote the norm of the magnetic ordering vector at the instability $Q$. For visualization purposes, we indicate the different magnetic instabilities by colored regions of the diagram. In panel $(b)$, we show the magnetic ordering vector $Q$ and the density of states as a function of doping. Panel $(c)$ shows the non-zero components of the magnetic order parameter at the instability $[\Delta_s]_{dd}$ for all orbitals $d$ for the same electron doping values.
}
\label{fig:phase_diagram}
\end{figure}

\subsection{Magnetic instabilities}\label{sec:rpa-chi}
Here, we analyze magnetism and superconductivity in the weak-to-intermediate coupling regime under the
matrix random-phase approximation. 
At low energies, weak-coupling RPA is justified by a parquet renormalization group analysis \cite{Fernandes2016}.
Within this framework, an infinite sum of single-scattering momentum $\boldsymbol{q}$ Feynman diagrams is performed exactly.
It has been shown that both RPA and vertex correction diagrams are taken into account in this approximation \cite{Altmeyer2016}.
Originating from spin fluctuations, magnetism is analyzed through the Stoner criteria, while superconductivity is studied using an eigenvalue equation for the electron-electron pairing strength as a function of the superconducting gap symmetry.

The bare susceptibility is given by
\begin{equation}
    \begin{split}
        \chi_{st}^{pq}(\boldsymbol{q},i\nu_m) = -\frac{T}{N_{\boldsymbol{k}}}\sum_{\boldsymbol{k}}\sum_{i\omega_n} G_{sp}(\boldsymbol{k},i\omega_n)G_{qt}(\boldsymbol{k}+\boldsymbol{q},i\omega_n+i\nu_m),
    \end{split}
\label{eq:chi0}
\end{equation}
where $i\omega_n$ ($i\nu_m$) ar fermionic (bosonic) Matsubara frequencies at temperature $T$, the momentum summation is performed over $N_{\boldsymbol{k}}$ $\boldsymbol{k}$-points, and
\begin{equation}
    G_{sp}(\boldsymbol{k},i\omega_n) = \sum_{\mu} \frac{ a_\mu^s(\boldsymbol{k})a_\mu^{p*}(\boldsymbol{k}) }{i\omega_n-E_\mu(\boldsymbol{k})} \label{eq:GF}
\end{equation}
is the non-interacting Green's function for the Hamiltonian $H_{\text{TB}}$ with eigenvalue $E_\mu(\boldsymbol{k})$ and eigenvector elements $a_\mu^s(\boldsymbol{k})$.
Here, we perform the Matsubara summation of Eq.~\eqref{eq:chi0} using Ozaki method \cite{Ozaki2007}.

The RPA spin and charge susceptibilities are given by
\begin{align}
    \hat{\chi}_s(\boldsymbol{q},i\nu_m) &= \hat{\chi}_0(\boldsymbol{q},i\nu_m)\left[ \hat{1} - \hat{U}_s\hat{\chi}_0(\boldsymbol{q},i\nu_m) \right]^{-1} \label{eq:chis} \\
    \hat{\chi}_c(\boldsymbol{q},i\nu_m) &= \hat{\chi}_0(\boldsymbol{q},i\nu_m)\left[ \hat{1} + \hat{U}_c\hat{\chi}_0(\boldsymbol{q},i\nu_m) \right]^{-1}, \label{eq:chic}
\end{align}
where the interaction matrices $\hat{U}_{s/c}$ have matrix elements \cite{Graser2009}
\begin{equation}
    \begin{split}
        & (U_s)^{dd}_{dd}=U, \; (U_s)^{dd}_{dd}=U', \;(U_s)^{dq}_{dq}=(U_s)^{dq}_{qd}=J, \\
        & (U_c)^{dd}_{dd}=U, \; (U_c)^{dd}_{qq}=-U'+2J, \;(U_c)^{dq}_{qd}=J, \; (U_c)^{dq}_{dq}=-J+2U'. \\
    \end{split}
\label{eq:U_matrices}
\end{equation}
The divergence of the susceptibility denominators imply the eigenvalue equations
\begin{align}
    \alpha_s\hat{1} - \hat{U}_s\hat{\chi}_0(\boldsymbol{q},0) &= 0 \label{eq:stoner_s} \\
    \alpha_c\hat{1} + \hat{U}_c\hat{\chi}_0(\boldsymbol{q},0) &= 0. \label{eq:stoner_c}
\end{align}
When the eigenvalue $\alpha_{s/c}=\pm1$, the susceptibility diverges marking the onset of a magnetic or charge instability.
The eigenvector $\bar{\boldsymbol{\Delta}}_{s/c}$ corresponding to the leading eigenvalue of Eqs.~\eqref{eq:chis} and \eqref{eq:chic} carries information on the order parameter at the instability \cite{Boehnke2015}.
Reminding that $J=U/4$ in the present analysis, the onsite Hubbard $U$ is the only free variable so that we solve Eqs.~\eqref{eq:stoner_s} and \eqref{eq:stoner_c} for eigenvalues $\tilde{\alpha}_{s/c}/U^c_{s/c}\equiv\alpha_{s/c}/U$ and simply impose $\tilde{\alpha}_{s/c}=1$ to obtain the critical Hubbard interaction $U^c_{s/c}$, which is used to probe for the susceptibilities close to the instability.
Finally, the magnetic ordering vector $\boldsymbol{Q}$ is obtained by the leading eigenvalue of Eqs.~\eqref{eq:chis} and \eqref{eq:chic} for all $\boldsymbol{q}$ within the Brillouin zone.
Given the magnetic ordering vector $\boldsymbol{Q}$ and the order parameter $\bar{\boldsymbol{\Delta}}_{s/c}$, the spin density texture in real space has its periodicity given by the magnetic wave-vector
\begin{equation}
    [\bar{\Delta}_s]_{dd}(\boldsymbol{r})\equiv[\bar{\Delta}_s]_{dd}e^{-i\boldsymbol{Q}\cdot\boldsymbol{r}}.
\label{eq:spin_texture}
\end{equation}
This expression can be simply seen as a Fourier transformation that selects the nesting vector $\boldsymbol{Q}$ as having the main weight in momentum space.
Finally, in Appendix \ref{sec:sc} we show the superconductivity methodology in more detail. We found superconducting gap symmetries as a function of electron doping ranging from the waves $d_{x^2-y^2}$ to $d_{xy}$ and then a product of both of them, $g_{xy(x^2-y^2)}$.

\section{Results}
\label{sec:results}

The electron-doping $n$ phase diagram of the present model for LaCrOAs is shown in Fig.~\ref{fig:phase_diagram}$a)$, which presents the spin critical Hubbard parameter $U_s^c$ obtained by solving Eq.~\eqref{eq:stoner_s}.
The spin fluctuations were found to always be more relevant than the charge fluctuations, with the condition $U_s^c<U_c^c$ always numerically obtained.
The color code in the figure indicates the norm of the nesting vector $Q=|\boldsymbol{Q}|$ found for each point in the phase diagram in the folded Brillouin zone view.
As a function of doping, the nesting vectors change abruptly revealing a magnetic phase diagram that starts with an AF state, transitioning to an AF stripe, and two distinct SDW states, which we differ by calling them incommensurate AF and incommensurate AF stripe.
The critical Hubbard interaction changes very little with doping but still, a plethora of magnetic ordering vectors reveals a strongly doping-dependent magnetic phase diagram.

The norm of nesting vectors $Q$ is presented in Fig.~\ref{fig:phase_diagram}$b)$ along with the density of states profile as a function of electron doping $n$.
The peaks in the DOS, caused by Lifshitz transitions, coincide with the two major changes in the nesting vectors.
These changes have drastic consequences to the orbital contribution within the distinct magnetic states.
The matrix elements $[\bar{\Delta}_s]_{dd}$ of the magnetic order parameter at the instability that is, when $U=U_s^c$, are shown as a function of doping $n$ in Fig.~\ref{fig:phase_diagram}$c)$.
We note that only the homogeneous elements $d=q$ are non-zero.
Colors indicate the contribution of each orbital
for $[\bar{\Delta}_s]_{dd}$, where $d=\{d_{xy},d_{x^2-y^2},d_{yz},d_{xz},d_{3z^2-r^2}\}$.
At the the low-doping region $n<4.25$, the main contribution to the order parameter comes from the $d_{3z^2-r^2}$ orbital.
In fact, the nesting vectors shown in Fig.~\ref{fig:orbital}$a,b)$ reveal that nesting connects two FS slices of the same $d_{3z^2-r^2}$ orbital.
On the other hand, Fig.~\ref{fig:phase_diagram}$b)$ also shows that at $n\approx4.25$ there is a doping region wherein the $d_{3z^2-r^2}$ and $d_{xy}$ orbitals closely interplay, and then the $d_{xy}$ orbital becomes more important for the magnetic order parameter at further doping.
The second Lifshitz transition at $n\approx4.4$, as shown in Fig.~\ref{fig:orbital}$c,d)$, in fact introduces two new pockets with dominating $d_{xy}$ contribution and which feature the magnetic nesting.

\begin{figure}
\includegraphics[width=1.0\columnwidth]{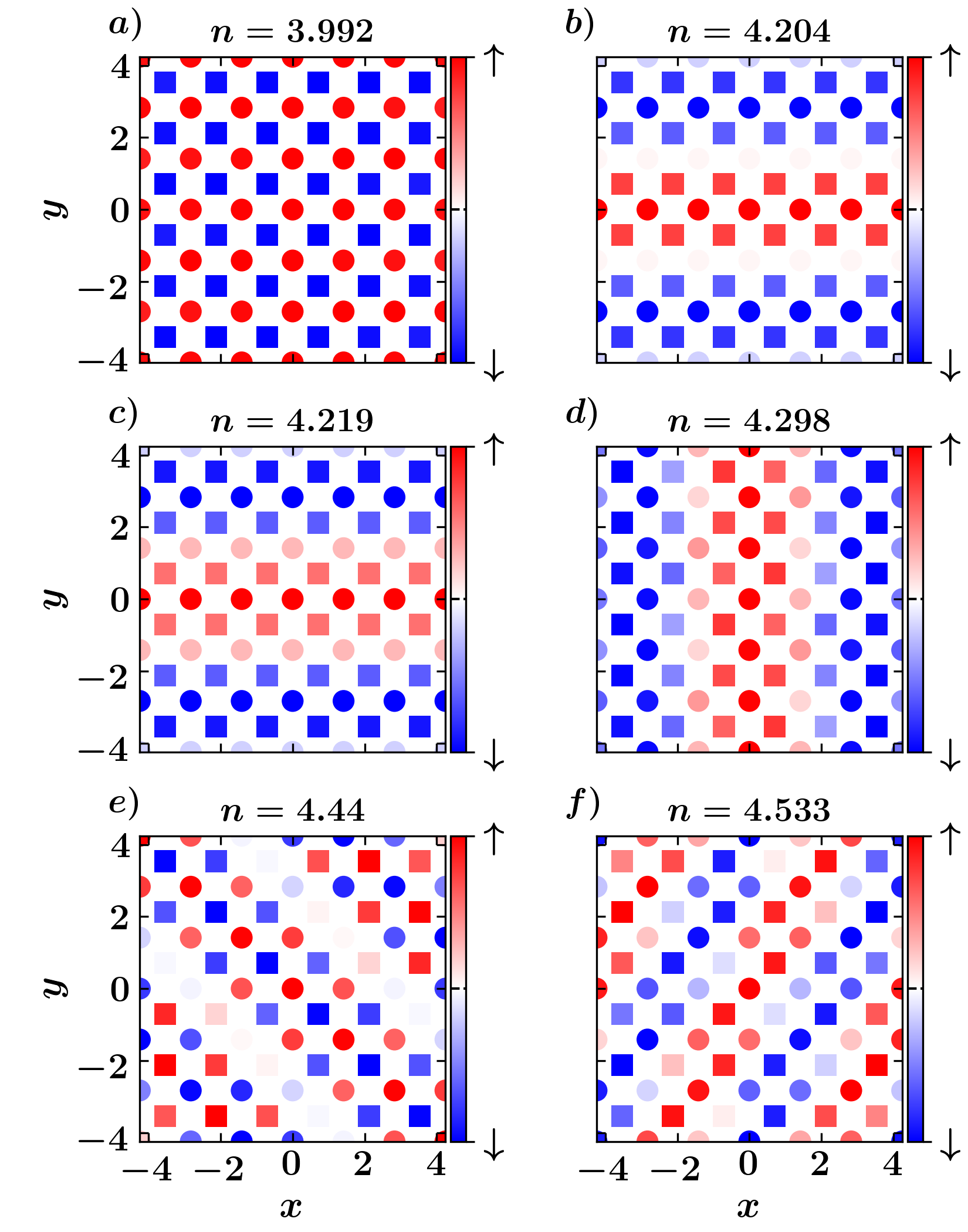}
	\caption{Spin density $[\bar{\Delta}_s]_{pp}(\boldsymbol{r})$ mapping the magnetic texture (spin-up: red; spin-down: blue) at Cr sites (Site A: circle; Site B: square) under varying electron doping concentrations: $n=3.992$ (panel a), $n=4.204$ (panel b), $n=4.219$ (panel c), $n=4.298$ (panel d), $n=4.440$ (panel e), and $n=4.533$ (panel f).}
\label{fig:magnetism}
\end{figure}

In Fig.~\ref{fig:magnetism}, we show the various magnetic patterns in the unfolded unit cell by computing Eq.~\eqref{eq:spin_texture} along the phase diagram of Fig.~\ref{fig:phase_diagram}$a)$.
In panel $a)$, the AF state is clear in the unfolded unit cell, while panels $b$-$d)$ show the emergence of the AF stripe.
Panel $e)$ shows an AF pattern along the $(\hat{x}-\hat{y})$ directions of the square lattice, while an incommensurability breaks the periodicity along the $(\hat{x}+\hat{y})$ direction.
The AF pattern along one direction and incommensurability along the other directions motivates labeling this SDW state as an incommensurate AF instability.
Finally, Fig.~\ref{fig:magnetism}$f)$ shows a pattern that resembles the stripe AF of panels $b$-$d)$ but incommensurate, producing wiggly stripes, which motivates labeling the instability incommensurate stripe AF.

\section{Discussions}
\label{sec:discussions}

Our findings provide a clear path to understanding the plethora of magnetic phases in RCrAsO, where R is a rare earth, as rooted in both Fermi surface topology and orbital specific properties of the eletronic states in the vicinity of the Fermi level.
This is indeed a general feature of the Fe-SC as suggested by ARPES measurements of Cr- and Mn-substituted BaFe$_2$As$_2$ showing that doping of electronic states in vicinity of the Fermi level alone does not provide an explanation for the magnetic phase diagram of these materials \cite{Cantarino2023,Cantarino2024}.

Our matrix random-phase approximation calculations based on a DFT-derived tight-binding model shows that charge doping changes the main orbital contribution for the magnetic state of RCrAsO from $d_{3z^2-r^2}$ to $d_{xy}$.
This trend in the magnetic phase diagram is connected to the orbital contribution to the Fermi surface, where, upon doping, the emergence of a new Fermi surface pocket enhances the $d_{xy}$ contribution.
The Fermi surface topology is the main responsible for the magnetic ordering vector $\boldsymbol{Q}$, which dictates the periodicity of the magnetic texture.
In the low doping region, an AF state is predicted, which connects with the dominant $d_{3z^2-r^2}$ orbital contribution.
With doping, the $d_{3z^2-r^2}$ loses weights at the Fermi surface and the $d_{xy}$ orbital starts to play a more important role.
This intermediate region features a stripe AF state.
With further doping, the $d_{xy}$ orbital dominates, with the appearence of a new Fermi surface sheet, and an incommensurate AF state is follwed by incommensurate stripe orders.


The possible overlap between orbitals of neighbor Cr sites in the extended unit cell
is favored by the $xy$-plane $d_{xy}$ electrons ~\cite{Chen2015, Lou2022, Bascones2012}. In contrast, $z$-axis magnetism of the
$d_{3z^2-r^2}$ electrons tends to localize the magnetic spins in the Cr sites. The localized
$d_{3z^2-r^2}$ are then connected with the commensurate AF order, which is perturbed by
the overlapping in-plane $d_{xy}$ orbitals which, eventually, break the N\'eel-type AF
and favor a stripe AF, and further incommensurate orders. Thus, a low-doping
localized magnetic picture is suppressed by the electron-doped itinerant-electron
magnetism, as proposed in Ref.~\cite{Bascones2012} for iron pnictides.

Electron doping drives a rich landscape to tune magnetism in the Cr-based materials, since electron doping changes Fermi surface properties and the main orbital character of the states contributing to the magnetic order parameter.
In this direction, adopting BaCr$_2$As$_2$ as reference compunds, it can also be interpreted that electron doping tunes the magnetism from a localized AF Neél-type phase to an itinerant stripe-type SDW phase in BaFe$_2$As$_2$.
In terms of the orbital character of the electronic states, Cr substitutions tune the $d_{3z^2-r^2}$ closer to the Fermi level \cite{Cantarino2023,Nayak2017,Richard2017}.

We note that the change in orbital hybridizations and the respective symmetries of electronic states in the vicinity of the Fermi surface can be indeed a consequence of charge, hole or electron, doping but not exclusively, as shown for Co- and Mn-substituted BaFe$_2$As$_2$ \cite{Figueiredo2022}.
Still, we conjecture that our finding of different magnetic phases associated with each orbital contribution, $d_{3z^2-r^2}$ or $d_{xy}$, is a general result even for a mechanism that is not charge doping.
Moreover, our work supports attempts of phase diagrams similar to BaCr$_2$As$_2$ such as Ba$[$Cr$_{(1-x)}$Fe$_x$$]_2$Cr$_x$As$_2$ in search for incommensurate magnetic phases.


\section*{CRediT authorship contribution statement}
\textbf{Henri G. Mendonça:} Formal Analysis, Visualization, Writing - review and editing.
\textbf{George B. Martins:} Conceptualization, Writing - review and editing, Formal analysis, Software
\textbf{Lauro B. Braz:} Conceptualization, Writing - review and editing, Formal analysis, Software.

\section*{Declaration of competing interest}
The authors declare that they have no known competing financial interests or personal relationships that could have appeared to influence the work reported in this paper.

\section*{Acknowledgements}
The authors are grateful to Maria Calder\'on, Leni Bascones, Wan-Sheng Wang, Qiang-Hua Wang, and Andrey Chubukov
for fruitful discussions on the subject matter of this work. We specially thank Fernando Garcia for a critical review of this work.
LBB acknowledges financial support from the S\~ao Paulo Research Foundation (FAPESP), Brazil (process number 2023/14902-8), and HPC resources provided by the Superintendency of Information Technology at the University of S\~ao Paulo.
HGM acknowledges financial support from Coordenação de Aperfeiçoamento de Pessoal de Nível Superior (CAPES) and HPC resources provided by the CENAPAD-SP.

\appendix 

\section{$d$-wave superconductivity}
\label{sec:sc}
We probe superconductivity by diagonalizing the Fermi surface-projected integral eigenvalue equation \cite{Graser2009}
\begin{equation}
    \int_{\text{FS}} d\boldsymbol{k}' \Gamma(\boldsymbol{k},\boldsymbol{k}')\Delta_{\alpha}(\boldsymbol{k}')=\lambda_{\alpha}\Delta_{\alpha}(\boldsymbol{k}),
\label{eq:eigenvalues}
\end{equation}
where $\Gamma(\boldsymbol{k},\boldsymbol{k}')$ is the kernel function and $\lambda_{\alpha}$ (eigenvalues) and $\Delta_{\alpha}(\boldsymbol{k})$ (eigenvectors) are the pairing strength and normalized gap function with symmetry $\alpha$.
For LaCrOAs, we found symmetries $\alpha=d_{x^2-y^2},d_{xy}$, and $g_{xy(x^2+y^2)}$.

According to the matrix-random-phase approximation results, superconductivity in the Cr pnictides could emerge with doping if the Hubbard $U$ is weak coupling.
We show in Fig.~\ref{fig:phase_diagram}(a) two horizontal lines demonstrating two possible $U$ values, for which superconductivity from spin fluctuations can emerge when the $U<U_s^c$ condition is matched.
For $U<0.6$ eV, spin fluctuation-driven superconductivity can emerge for the doped compounds, while for $0.65<U<0.6$, a double-dome structure is predicted.

\begin{figure}
\centering
\includegraphics[width=\columnwidth]{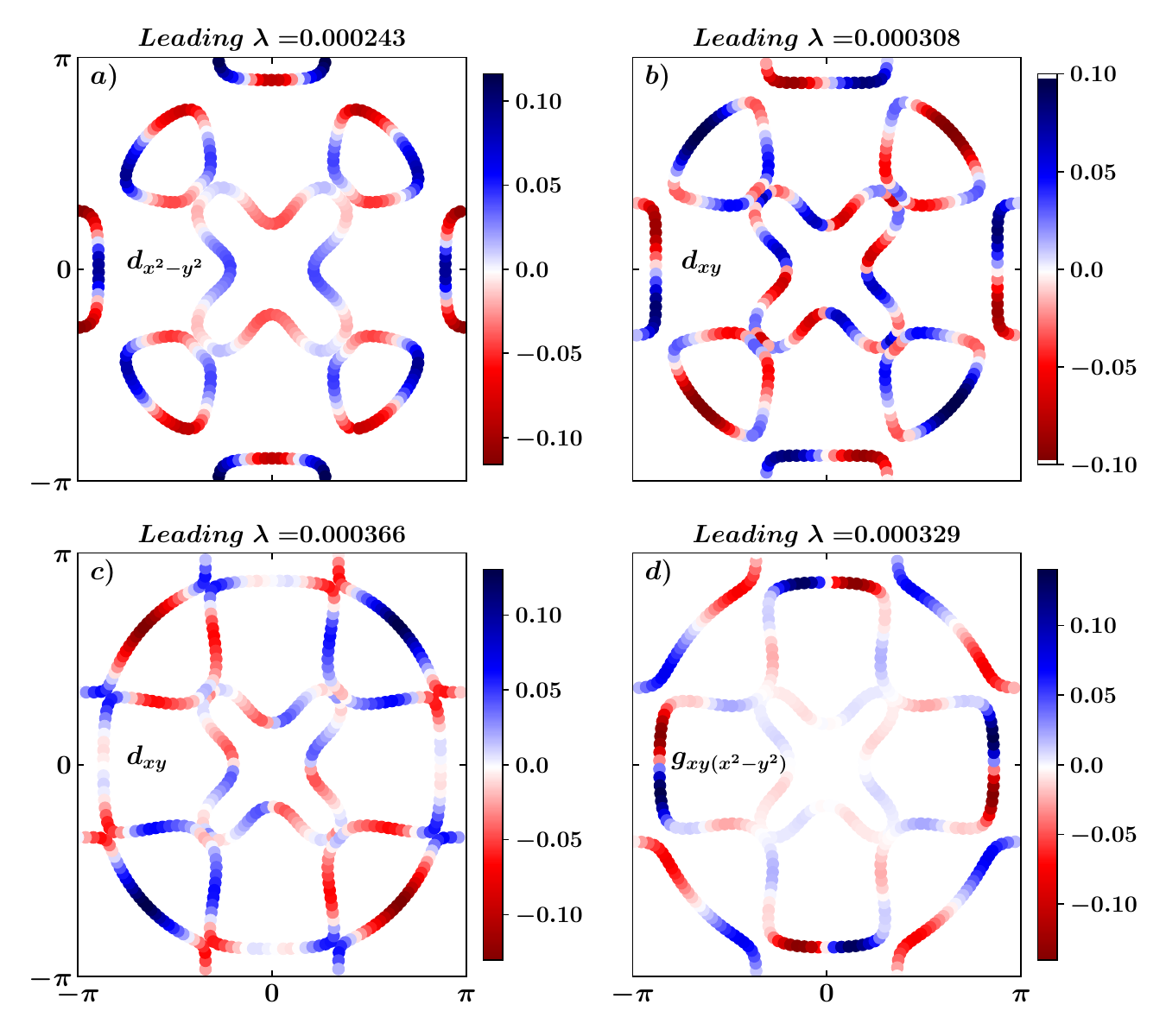}
	\caption{ Superconducting symmetries for different doping levels, $n = 4.17, 4.30, 4.44, 4.53$ respectively in panels (a) to (d). 
}
\label{fig:sc}
\end{figure}

Fig.~\ref{fig:sc} shows the different superconducting gap symmetries found by solving Eq.~\eqref{eq:eigenvalues} for different filling values.
At low doping, the $d_{x^2-y^2}$ wave is dominating [panel (a)].
However, this symmetry moves to a $d_{xy}$ wave for mid-$n$ values [panels (b) and (c)] and finally, becomes a $g_{xy(x^2-y^2)}$ wave [panel (d)] at the high-doping regime.

\bibliographystyle{elsarticle-num}
\bibliography{lacr}

\begin{thebibliography}{10}
\bibitem{note1}
Note that in this case the cuprates are being hole-doped and the Fe-SC are
  being electron-doped.
\end{thebibliography}

\end{document}